\documentstyle[11pt,newpasp,twoside,epsf]{article}
\def\edcomment#1{\iffalse\marginpar{\raggedright\sl#1\/}\else\relax\fi}
\reversemarginpar

\newcommand{\lbol}{L$_{bol}$}
\newcommand{\llyc}{L$_{LyC}$}
\newcommand{\lk}{L$_{K}$}
\newcommand{\nsn}{$\nu_{SN}$}

\newcommand{\apjj}[2]{Ap. J., #1, #2.}

\newcommand{\apjjl}[2]{Ap. J. (Letters), #1, #2.}

\newcommand{\ajj}[2]{A. J., #1, #2.}

\newcommand{\vol}[1]{1}

\newcommand{\solm}{M$_{\odot}$}

\newcommand{\rf}{\par\noindent\hangindent 15pt {}}

\begin{document}
\title{Stellar Populations in the Nucleus of NGC~6764}
 \author{Andreas Eckart}
\affil{
Universit\"at zu K\"oln, I.Physikalisches Institut, Z\"ulpicherstra\ss e 14
\\
50937 K\"oln, Germany
\\
and
\\
MPI f\"ur extraterrestrische Physik, 88740 Garching, Germany,
}
\author{Eva Schinnerer}
\affil{Astronomy Department, California Institute of Technology, Pasadena,
CA~91101, USA}

\begin{abstract}
We present first results of near-infrared integral field spectroscopy
of the central 8''$\times$8'' of the Wolf-Rayet nucleus of the barred spiral
galaxy
NGC~6764. In addition to stellar CO and Na~I absorption lines as well as
recombination lines of H and He~I the K-band spectrum shows
strong emission from molecular hydrogen.
Analysis of the stellar nuclear light using population synthesis in
conjunction with NIR spectral synthesis suggests either the presence of two
starbursts with ages of $\sim$ 5 Myrs and $\sim$ 20 Myrs or continuous
star formation with an star formation rate of
$\sim$ 0.4 \solm yr$^{-1}$ ~over the past 1~Gyr.
\end{abstract}


\section{INTRODUCTION}

NGC~6764 is a nearby (32~Mpc for $H_0$=75 km s$^{-1}$; 
1$^{\prime\prime}$=160~pc)
S-shaped barred spiral galaxy (SBb), classified as
a LINER galaxy on the basis of optical spectroscopy.
The galaxy contains a nuclear stellar optical 
continuum source with a ``width" of $\sim 1.6"$ or about 200~pc
(Rubin, Thonnard, \& Ford 1975).
NGC~6764 is unusual in that it displays
a prominent 466 nm Wolf-Rayet emission feature
at the nucleus.
Wolf-Rayet emission features are rarely observed in galaxies, and
their presence is indicative of very recent massive star formation
(Armus, Heckman \&  Miley 1988; Conti 1991).
The study of such objects is 
essential for understanding the starburst phenomenon in galaxies. 
\\
A first analysis of the starburst activity in NGC~6764 was presented in
Eckart et al. (1991) and Eckart et al. (1996) and 
revealed a dense concentration of molecular gas
and a very recent (at most a few times 10$^7$ years) starburst
at the nucleus of NGC~6764.
Here we present first K-band integral field spectroscopy
of the  NGC~6764 nucleus (see Fig.1).
Observational details as well as a more thorough
description of the analysis will be presented in a forthcoming paper.

\begin{figure}[htp]
\plotfiddle{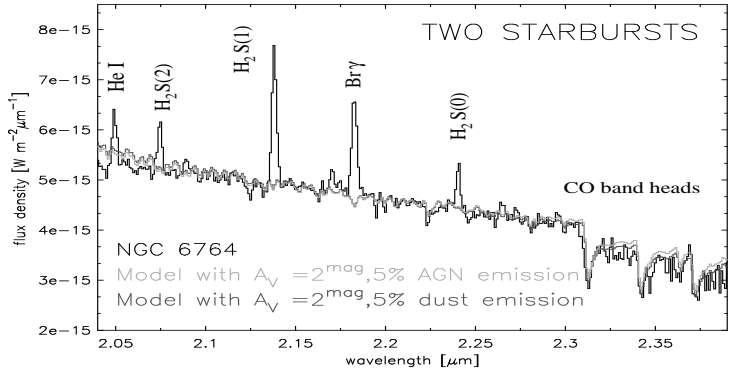}{4.0cm}{-90}{55}{40}{-160}{210}
\caption{NGC~6764 K-band spectrum (black) and the results of the
population and spectral synthesis modeling for two bursts (grey).}
\label{fig1}
\end{figure}

\section{THE NUCLEAR STAR FORMATION HISTORY}

To investigate the nuclear star formation history further, we use the 
population synthesis code STARS  which was
successfully applied to a number of galaxies (e.g. Krabbe,
Sternberg \& Genzel 1994, NGC~1808) in
conjunction with the NIR spectral synthesis code SPECSYN (Schinnerer et al.
1997 and references therein). 
 STARS has
as output observable parameters such as the bolometric luminosity
L$_{bol}$, the $K$ band luminosity L$_K$, the Lyman continuum luminosity 
L$_{LyC}$ and the supernova rate $\nu_{SN}$, as well as the diagnostic
ratios between these quantities:
L$_{bol}$/L$_{LyC}$, L$_K$/L$_{LyC}$ and 10$^{9}\nu_{SN}$/L$_{LyC}$.
All three ratios are measures of the time evolution and the shape of
the IMF, with slightly different dependencies on its slope $\alpha$ and
upper mass cutoff m$_u$.
H-R diagrams representing the distribution of these luminosities are 
calculated. SPECSYN uses the distribution of K band luminosity L$_K$
within the H-R diagram to weight standard star spectra of 
different spectral type and luminosity (Schinnerer et al. 1997,
Schinnerer et al. in prep).

\subsection{The Parameters for the Population Synthesis}

NIR integral field spectroscopy revealed that the star
formation is concentrated on the nuclear region.
To estimate {\it \lbol} for the central 3'' ($\sim$ 500~pc) of NGC~6764 
we followed the approach of Eckart et al. (1991).
{\it \lk} and {\it \llyc} were estimated via the equations given in
Genzel et al. (1995) (using a $\delta \lambda$ = 0.6$\mu$m as the $K$
band width) and taking the values measured with 3D in a 3'' aperture. 
From our analysis we infer an extinction towards the nuclear stellar
cluster of about (2 - 3)$^{mag}$ in agreement with the observed
Br$\delta$/Br$\gamma$ ratio and earlier findings by Eckart et al.
(1996). In addition we see evidence for a 5\% non-stellar contribution
to the $K$ band continuum either from warm dust or
a power law contribution from the AGN itself.
For {\it \nsn} we used the relation given by Condon
(1992) and the nuclear 5 GHz flux of 3.4 mJy from Baum et al.
(1993).

\begin{figure}[htp]
\plotfiddle{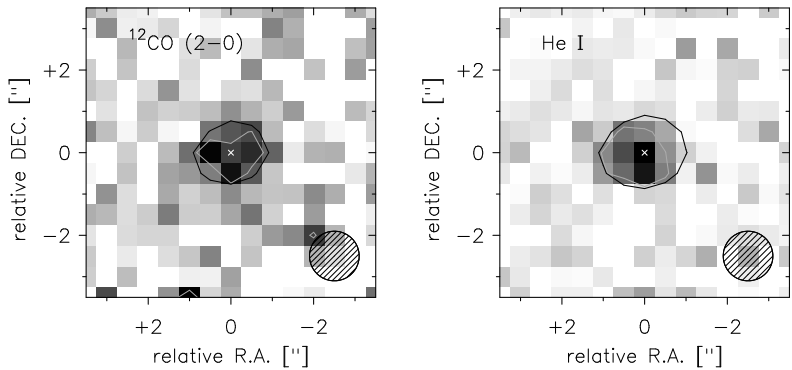}{6.0cm}{-90}{55}{55}{-200}{300}
\caption{Maps of the 2.29$\mu$m ~~$^{12}$CO 2-0 absorption (left) and
2.06$\mu$m ~~He~I emission line towards the nucleus of NGC~6764.
The black solid contour line is the 50/
continuum emission and the white solid line is the 3$\sigma$ contour
line of the line absorption and emission.
The beam in the lower left corner of the maps has a FWHM of 1.2''.}
\label{fig2}
\end{figure}

\subsection{Results of the Modeling}

Given the presence of the WR stars two scenarios have been considered:
(1) Two starburst events and (2) continuous star formation.
Both possibilities are explored in the following.
\\
\\
{\it Two Young Starbursts:}
The Wolf-Rayet features indicate a very recent star formation
event only about 5 Myrs ago.
This immediately implies a second starburst event to account for the
presence of the cool evolved stars indicated by the stellar
absorption features (see fit in Fig.1).
The current appearance of the NGC~6764 nucleus
may therefore be dominated by two major starburst events:
\\
\\
{\it Starburst about 5~Myrs ago (SB\#1):}
Assuming a starburst event with a decay time of 3~Myrs 
seems reasonable given the fact that SN explosions might disturb the ISM
medium and prevent further star formation . For a 
starburst of that age STARS gives a mean hot star of type O5 with a
mean effective temperature of T$_{eff}$=45000K. Boer \& Schulz (1989)
find an electron density of of $n_e$ $\approx$ 6$\times$10$^2$cm$^{-3}$
(with $T_e$=10$^4$K). Neglecting dilution effects
this gives an HeI/Br$\gamma$ ratio of 0.7 (Lan\c{c}on
\& Rocca-Volmerage 1996). Comparison of $K$ band spectra from active
galaxies with an AGN
and starburst galaxies (Vanzi et al. 1998) suggests that strong HeI line
emission is only present in starburst galaxies.
This may indicate that most of the HeI line flux observed towards the
nucleus of NGC~6764 is due to young, hot stars and not due to an AGN
component.
This allows us to correct the observed HeI/Br$\gamma$ ratio and obtain
the Br$\gamma$ line flux that is associated with this starburst event. We
derive that about 65\% of the Br$\gamma$ line flux is due to an 5~Myrs
old starburst.
\\
\\
{\it Starburst about 16 Myrs ago (SB\#2):}
This second starburst has to account for about 94\% of the total \lk ~and the
total amount of \nsn. If the age is about 20~Myrs all 
Br$\gamma$ line flux not accounted by SB\#1
could be associated with the LINER nucleus.
For starbursts with ages of  $\sim$ 1~Gyr AGB stars have a prominent
contribution to the $K$ band continuum
(Lan\c{c}on \& Rocca-Volmerage 1994,
Lan\c{c}on \& Rocca-Volmerage 1996).
However, in the case of the NGC~6764 nucleus
a starburst of this age cannot account for the observed extended
nonthermal radio emission which is very likely due to
star formation, e.g. SN explosions.
A starburst age of $\sim$ 1~Gyr therefore does
not appear to be very  plausible for SB\#2.
This scenario is similar to the one found in IC~342 where a 70~pc
diameter starburst ring with an age of about 5~Myrs surrounds a
nuclear starburst of about 15~Myrs (B\"oker, F\"orster-Schreiber \& Genzel 
1997). 
\\
\\
{\it Continuous Star Formation:}
Assuming continuous star formation in the nuclear region the stellar
cluster must have an age of $\leq$ 1 Gyr, since enough cool evolved
stars are produced to obtain the observed ratio of L$_K$ to L$_{LyC}$
ratio. It is necessary to reach the AGB phase to obtain a reasonable
fit, since only then the cool luminous stars are numerous. This
analysis is, however, hampered by the fact, that the
currently available evolutionary tracks of the high mass stars
in general do not produce very cold red super-giants
(e.g. of type M) and that SPECSYN does not use spectra of AGB stars.
We use evolutionary tracks extended for the AGB phase (N.M. Schreiber 1998,
PhD thesis, LMU Munich)
and approximate the spectra of AGB stars by those of a red M type
super-giant (RSG). In this scenario, it is
expected that the WR emission and the stellar absorption bands would
have similar distribution even at highest spatial resolution.
The corresponding fit to the data is of similar quality as the fit for two
bursts shown in Fig.1.
\\
\\
{\bf In summary:} The hot stars
(HeI/Br$\gamma$ ratio of 0.46$\pm$ 0.06 indicating
T$_{eff}$ $\sim$ 35 000 K),
the Wolf-Rayet stars
(466 nm Wolf-Rayet emission feature)
and the evolved cool stars
(see CO absorption band heads at $\lambda \ge$ 2.29 $\mu$m and NaI
absorption at 2.21 $\mu$m)
could be either co-spatial or might form a young starburst ring around the
older nuclear stellar cluster as observed in IC~342 (B\"oker,
F\"orster-Schreiber \& Genzel 1997).
Our analysis indicates two possible nuclear star formation histories:
(1) Two starbursts with ages of about 5~Myrs and about 16~Myrs and
decay times of 3~Myrs producing the WR stars and red super-giants that
contribute to a large amount of the $K$ band continuum.
(2) Continuous star formation with a SFR of $\sim$ 0.4 \solm/yr for at
least 1~Gyr.
\\
\\
\footnotesize
\rf{Armus, L., Heckmann, T.M., Miley, G.K., 1988, \apjjl{326}{L45}}
\rf{Baum, S.A., O'Dea, C. P., Dallacassa, D., de Bruyn, A. G., Pedlar, A.,
    \apjj{419}{553}}
\rf{Boehr, B. and Schulz, H., 1990, {\em Astrophs. Space Sci. }{\bf 163},201.}
\rf{B\"oker, T., F\"orster-Schreiber, N.M., Genzel, R., 1997,
    \ajj {114}{1883}}
\rf{Condon, J.J., 1992, Ann.Rev.Astr.Ap. 30, 575}
\rf{Conti, P.S., 1991, \apjj{377}{115}}
\rf{Eckart, A., Cameron, M., Boller, Th., Krabbe, A., Blietz, M.,
    Nakai, N., Wagner, S.J., Sternberg, A., 1996, \apjj{472}{588}}
\rf{Eckart, A., Cameron, M., Jackson, J.M., Genzel, R., Harris, A.I.,
    Wild, W., Zinnecker, H., 1991, \apjj{372}{67}}
\rf{Genzel, R., Weitzel, L., Tacconi-Garman, Blietz, M., Krabbe, A.,
    Lutz, D., Sternberg, A., 1995, \apjj{444}{129}}
\rf{Krabbe, A., Sternberg, A., and Genzel, R., 1994, \apjj{425}{72}}
\rf{Lan\c{c}on, A., Rocca-Volmerange, B., 1996, New. A., 1, 215}
\rf{Lan\c{c}on, A., Rocca-Volmerange, B., 1994, Ap\&SS 217 271}
\rf{Rubin, V.C., Thonnard, N., Ford, W.K.,1975, \apjj{199}{31}}
\rf{Schinnerer, E., Eckart, A., Quirrenbach, A., B\"oker, T., Tacconi-Garman, 
    L.E., Krabbe, A., Sternberg, A., 1997, \apjj{488}{174}}
\rf{Schinnerer, E., Eckart, A., Tacconi, L.J., 1998, \apjj{500}{147}}
\rf{Vanzi, L., Alonso-Herrero, A., Rieke, G. H., 1998, \apjj{504}{93}}

\end{document}